\documentclass{article}
\usepackage{amssymb}
\usepackage{graphicx}
\usepackage{amsmath}
\usepackage{epsfig}
\def\Red{}
\def\Black{}
\def\Blue{}

\begin{document}

\title{
\Red
Neutrino masses and non abelian horizontal symmetries
\Black
}
\author{V.~Antonelli, F.~Caravaglios, R.~Ferrari, M.~Picariello \\
Universit\`{a} di Milano, and INFN,  sezione di Milano}
\date{October 2, 2002}
\maketitle

\begin{abstract}
\Blue
Recently neutrino experiments have  made  very significant  progresses
and our knowledge  of neutrino masses and mixing has considerably improved.
 In a model
independent Monte Carlo approach,  we have examined a very large class of 
textures, in the context of non abelian horizontal symmetries;
 we have found that neutrino data select    only  those charged lepton matrices
with left-right  asymmetric texture. The large atmospheric 
mixing angle needs   $m_{23}\simeq m_{33}$. This result,
if combined with similar recent  findings  for the quark sector in the $B$
oscillations, can be interpreted as a hint for  $SU(5)$ unification. 
In the neutrino sector  strict 
neutrino  anarchy is disfavored by data, and at least a factor 2 of 
suppression in the first row and column of the neutrino majorana mass matrix 
is required. 
\Black
\end{abstract}

\section{Introduction}

\noindent Understanding the origin of particle masses is one of the main
problems in particle physics. While strong and electroweak interactions are
described by a simple lagrangian arising from local gauge symmetries, the
sector that breaks these symmetries, giving mass to all existing particles,
is rather complicated and not yet understood. The electroweak boson masses
are very well described by the Higgs mechanism: if the scalar Higgs particle
is a $SU(2)_{\text{weak}}$-doublet, than the breaking induced by the vev of
such field would inevitably lead to a well known relation between masses and
the weak mixing angle 
\begin{equation}
\frac{M_{W}^{2}}{M_{Z}^{2}}=\cos ^{2}\theta _{w}.    \label{eq:1}
\end{equation}
The experimental success of the above relation  is an indirect prove that
the mechanism giving rise to particle masses is the spontaneous symmetry
breaking  of the $SU(2)$ symmetry through a scalar $SU(2)$ doublet. We
stress that the (\ref{eq:1}) is due to the spontaneous breaking of a $SO(4)$
 approximate symmetry into a custodial $SU(2)_{V}$ symmetry.
In other words, even if it is not ruled out that the Higgs particle is a
composite particle, its quantum numbers have been established. The same
general consensus is far from being achieved in the fermion sector, where
the large number of free parameters ( 33 mass matrix entries, including 6
free entries  for the majorana  neutrino mass matrix),  poses several
difficulties in disentangling the underlying symmetries at the origin of the
fermion mass spectrum. In a recent paper \cite{stocchi}, it has been
proposed a model independent approach to try to extract  from all available
measurements the hierarchy texture of quark mass matrices. Here we apply a
similar analysis in the lepton sector.

\subsection*{Experimental limits for neutrino masses}

The long standing problem of proving that neutrinos are massive particles,
determining the values of their masses and finding a natural explanation of
their lightness has been faced in many different ways during the last
century. For a review on this subject see, for 
instance~\cite{reviews,altarev,Frigerio}.

The first limit on neutrino mass has been obtained in 1947~\cite{primamassa}
by using the \emph{Fermi-Perrin} method based on the study of the $\beta$
spectrum near the end-point. The updated results of this kind of kinematical
searches~\cite{nutau}, together with the limits coming from the
experiments on the neutrinoless double $\beta$
 decay~\cite{Klapdor-Kleingrothaus:yx,Aalseth:2002rf},
 give the following upper limits
for tauonic and muonic neutrino masses: $m_{\nu_\tau} < 18.2 \, MeV$, $%
m_{\nu_\mu}<190 \, KeV$. The analogous result for the electron neutrino
comes from the \emph{Mainz} and \emph{Troitsk}
experiments~\cite{Bonn:tw,Lobashev:uu},  which pose the upper limit
 $m_{\nu_e} < 2.2 \, eV$.
In the next future a big improvement is expected by the study of the Tritium 
$\beta$ spectrum in the KATRIN~\cite{KATRIN} experiment, that is supposed to
improve the sensitivity down to $m_{\nu_e} < 0.35 \, eV$ at $90 \%$
confidence level.

The search for neutrinoless double $\beta$ decay is relevant also because,
if we assume CPT invariance, the existence of such a decay would imply the
Majorana nature of neutrinos~\cite{Pascoli:2001vr}. The best limit on this
process comes from the Heidelberg-Moscow
collaboration~\cite{Klapdor-Kleingrothaus:yx}
 $\langle m_{\nu} \rangle < 0.35 \, eV$ and from
IGEX  (International Germanium Experiment)~\cite{Aalseth:2002rf} $\langle
m_{\nu} \rangle < 0.33-1.35 \, eV$. In the last year there has been a claim~
\cite{Klapdor-Kleingrothaus:2001ke}  from some members of the
Heidelberg-Moscow collaboration of  discovery of a $2-3 \sigma$ effect that
would be a signal of  neutrinoless double $\beta$ decay with $\langle
m_{\nu} \rangle\approx 0.11 - 0.56 \, eV$.  A still open strong discussion,
 (see for example~\cite{Feruglio:2002af}),
 based mainly on the statistical analysis of the data has been related to
this discovery. One can say that at present  the importance generally given
to this ``effect'' has decreased, and the \emph{discovery} can be
interpreted more as a limit on $\langle m_{\nu} \rangle$.

Other indirect, but essential, information on neutrino masses are obtained
from the experiments looking for signals of neutrino oscillations, from
which one can extract the values of the mass differences between the
different mass eigenstates. These evidences come mainly from the study of
solar and atmospheric neutrinos, but also from the long baseline and the
LSND experiment.

The existence of a solar neutrino deficit has been proven both by the
radiochemical experiments~\cite{Homestake,SAGE,GALLEX,GNO} and by the more
recent ones using Cherenkov
detectors~\cite{Kamiokande,SKsolar,SNOCC,Ahmad:2002jz,Ahmad:2002ka}. After the
most recent SNO results~\cite{Ahmad:2002jz,Ahmad:2002ka}, we have a 
crystal clear proof that part of the electron neutrinos emitted by the Sun are
converted into other active flavors during their way to Earth. This
oscillation phenomenon is strictly connected to a value of neutrino mass
different from zero. For a detailed discussion of the impact of SNO data and
of the present values of the mass difference inferred from solar neutrino
experiments, we refer the interested reader to~\cite{AllNeutrino}. Here
we just recall that the most probable solution of the solar neutrino puzzle
is the so called LMA (Large Mixing Angle) solution, with best fit points for
the squared masses difference in the range $4-6 \times 10^{-5} \, eV^2$ and
for the mixing angle around $\tan^2 \theta_{sol} \simeq 0.4$.

Many robust evidences of oscillation came also by various atmospheric
neutrino experiments
\cite{Kamiokandeatm,SKatm,SKatmres,Becker-Szendy:vr,Sanchez:pj,MACRO}
and mainly by SuperKamiokande (SK) that is the one characterized by the higher
statistics. In these experiments the difference of the squared of the mass
eigenvalues responsible for the oscillation is higher than the difference
one finds in the solar neutrino case. The updated SK atmospheric results are~
\cite{SKatmres}, for instance: $\Delta m_{atm}^2 = (1.6 \simeq 3.9) \times
10^{-3} \, eV^2$ and $\sin^2 2 \theta_{atm} > 0.92$ at $90 \%$ of confidence
level.

The LSND experiment~\cite{Sung:ps} have found signals of 
$\nu_{\mu} (\bar{\nu}_{\mu}) \to \nu_{e} (\bar{\nu}_{e})$ oscillations, that 
would correspond to
very high values of $\Delta m^2$ (up to $\Delta m^2 \geq 1 eV^2$). This
result, if confirmed, could not be explained together with the ones about
solar and atmospheric neutrinos in the usual framework of three flavour
states. Hence, this has been one of the main motivations to introduce the
idea of the existence of at least one additional sterile neutrino. The LSND
results have not been confirmed by the KARMEN experiment~\cite{Wolf:2001gu},
but up to now they cannot be ruled out \footnote{About the
compatibility of LSND and Karmen results see
also~\cite{LSNDconKARMEN}}. Therefore there is a great expectation 
for the MiniBoone experiment~\cite{Hawker:pt},
that should produce data from 2004. It has been
built in such a way to test all the phase space of LSND results and confirm
or disprove them.

The long baseline accelerator experiment K2K~\cite{K2K} has confirmed, even
if with a statistic for the moment much lower than the one of atmospheric
and solar neutrino experiments, the existence of oscillations and it has
found values of the mixing-parameters in good agreement with the atmospheric
results. In future this and the other forthcoming European~\cite{CERNGSasso}
and American~\cite{longUSA} long baseline experiments should reach a much
higher statistic and they will be very important for the determination of
neutrino masses.

Every analysis of neutrino mass must also include the constraints coming
from the results of the reactor experiments CHOOZ~\cite{CHOOZ} and Palo
Verde~\cite{PaloVerde}. The absence of oscillation signals at these
experiments can be used to exclude a significant part of the
mixing-parameter plane.

\subsection*{Implementation of the experimental limits}

We have introduced in our analysis all the relevant experimental limits that
we have discussed in the previous subsection.

From the point of view of their impact on the final output, the most
significant experimental results are the ones concerning the mass
differences coming from different sources (atmospheric and solar neutrinos,
CHOOZ and the other reactor and accelerator experiments). The results for
the mass differences and the mixing angles coming from the different
categories of experiments have been treated as independent experimental
inputs and they all contributed to the determination of the total $\chi^2$
value.
Namely, to apply the experimental constraints  we have
defined the function 
\begin{equation}  \label{eq:chisquare}
\chi^2(O^{{\rm t}h})=\sum_{i}\left( \frac{O^{{\rm t}h}_i- O^{{\rm e}xp}_i}{%
\sigma^{{{\rm e}xp}}_i} \right)^2.
\end{equation}
where $O^{{\rm e}xp}_i\pm \sigma^{{\rm e}xp}_i$ are the experimental data. 
The function $\chi^2(O^{{\rm t}h})$ will be used
afterwords: by means of a weight $\exp( -\chi^2/2 )$, we will select models
with predictions $O^{{\rm t}h}$, close to the experimental data. Note that
for simplicity we have neglected correlations among different experimental
data.

We have also considered all the upper values of the masses for the different
neutrino flavors obtained by the direct kinematical searches, but their
impact to the results of our analysis is negligible.

Finally, we have inserted in the $\chi^2$ determination the constraint on
the electron neutrino mass that would come from the neutrinoless double $%
\beta$ decay, in case one assumes the value of the mass given
 by~\cite{Klapdor-Kleingrothaus:2001ke}. However, the impact of this additional
experimental constraint on the outputs of our analysis is marginal,
because the configurations that could satisfy this restriction would require
a very strong fine tuning of the phases and therefore they have a low
statistical significance (see also subsection 2.3.3). This fact can be, of 
course, considered also from a different point of view, in the sense that if 
the experimental debate on the evidence for a neutrinoless double beta decay 
would confirm this claim,
almost all the regions in our plots would be ruled out and we were left with
few points, corresponding to distinct solutions for neutrino mass matrix.

\section{Looking for model independent parameterizations}

Low energy lepton masses arise from the Yukawa interaction with the light
Higgs bosons: 
\begin{equation}
L_{Yuk}=Y_{l}^{ij}H_{d}\bar{L}_{i}l_{j}+\mathrm{h.c.}  
\label{eq:1bis}
\end{equation}
The Yukawa couplings $Y^{ij}$, unconstrained and incalculable in the
Standard Model, arise from more fundamental high energy Lagrangians. The
Lagrangian given in (\ref{eq:1bis}) is an effective one, where only
light particles appear as physical fields: at such low energy, heavy
particles can only appear as virtual internal propagators of Feynman
diagrams of the full theory. For example a process with $n$ light particles $%
\psi $ that goes into $k$ light particles $\phi $ through the virtual
exchange of  one heavy field $F$  (with mass $M$) through the interaction $%
g_{1}\psi ^{n}F+g_{2}\phi ^{k}F$ can be described by one single operator $%
g_{1}g_{2}\;\psi ^{n}\phi ^{k}/M^{2}$ containing only the light particles.
Similarly, the Yukawa interactions in equation (\ref{eq:1bis}) can arise
from higher dimensional operators $H_{d}\bar{L}_{i}l_{j}\phi ^{k}/M^{k}$ \
where the field $\phi $ acquires a \textit{vev }$\bar{\phi}=<\phi >$ and
thus $Y_{l}^{ij}=\bar{\phi}^{k}/M^{k}$. In general the heavy particles
content could be very complex,  but if  the original Lagrangian is
invariant under some (flavor) symmetry, the above mentioned  low energy
operators must obey the same symmetry. There are different  classes  of
symmetries that can explain the observed pattern of fermion masses and
mixing:  abelian \cite{abelian} and non abelian \cite{nonabelian} horizontal
symmetries, but also discrete symmetries \cite{discrete} have been studied 
in the past. Recently, neutrino masses and mixings have animated several 
discussions about models with extra dimensions \cite{extra}. A 
review of  theoretical ideas can be found in \cite{altarev}.

\subsection{U(2) horizontal symmetry}

An interesting class of models is based on a $U(2)$ horizontal symmetry. This
 symmetry is helpful to suppress potentially dangerous Flavor Changing Neutral
 Currents (FCNC) in supersymmetric models. 
It acts on the known fermion families as follows. The light
fermions transform as doublets $\mathbf{2}$ under the $U(2)$ group 
\begin{equation}
\left( 
\begin{array}{c}
e_{L} \\ 
\mu _{L}
\end{array}
\right) \left( 
\begin{array}{c}
\nu _{L}^{e} \\ 
\nu _{L}^{\mu }
\end{array}
\right) \left( 
\begin{array}{c}
e_{R}^{c} \\ 
\mu _{R}^{c}
\end{array}
\right)
\end{equation}

$f_{L}$ are the left handed $SU(2)_{weak}$ doublets, while $f_{R}^{c}$ are
the charge conjugated of the right handed $SU(2)_{weak}$ singlets. The light
Higgses (responsible for the electroweak breaking) are singlets as well as
the leptons of the third generation. The $U(2)$ group (differently from $SU(2)
$) includes a $U(1)$ phase transformation: the $\mathbf{2}$ and the $\mathbf{%
\bar{2}}$ are not equivalent representations; in particular such a $U(2)$
forbids Yukawa interactions like \cite{nonabelian} 
\begin{equation}
\mathrm{\ }g_{e}H_{d}\;\overline{e}_{L}e_{R}+g_{\mu }\mathrm{\ }H_{d}\;%
\overline{\mu }_{L}\mu _{R}
\end{equation}

as well as all possible mass or mixing terms concerning the two lightest
generations. On the contrary the $\tau $ can have mass since it is a singlet
under the above $U(2).$ To allow the lighter fermions acquiring a mass, we
need to break the $U(2)$ symmetry in two steps. Firstly, the breaking of $%
U(2)\rightarrow U(1)$ can be induced by a $U(2)$-doublet $\phi _{a}$, the $%
\mathbf{2}$, and a triplet $\Phi _{ab}$, the $\mathbf{3}$. Exploiting the $%
U(2)$ symmetry we can always rotate their $vev$'s in order to obtain%
\footnote{%
We label the two lightest families with 1 and 2.} $\langle \phi _{2}\rangle
=v_{1}$, $\langle \Phi _{22}\rangle =v_{2}$ and $\langle \phi _{1}\rangle
=\langle \Phi _{11}\rangle =\langle \Phi _{12}\rangle =\langle \Phi
_{21}\rangle =0$ which implies the following Yukawa couplings 
\begin{equation}
\mathrm{\ }gH_{d}\;\overline{\mu }_{L}\tau _{R}\;\phi _{2}+g^{\prime }%
\mathrm{\ }H_{d}\;\overline{\mu }_{R}\tau _{L}\;\phi _{2}^{\ast }+g^{\prime
\prime }\mathrm{\ }H_{d}\;\overline{\mu }_{L}\mu _{R}\;\Phi _{22}+\mathrm{%
h.c.}
\end{equation}
or in terms of the lepton mass matrix 
\begin{equation}
M_{lep}\propto \;\left( 
\begin{array}{ccc}
0 & 0 & 0 \\ 
0 & v_{2} & v_{1} \\ 
0 & v_{1} & M_{u}
\end{array}
\right)   \label{eq:14}
\end{equation}
At lower energy also the $U(1)$ can be broken by a $U(2)$-singlet $A_{ab}$,
anti symmetric under the exchange of the indices $a$ and $b$. This changes
the matrix (\ref{eq:14}) into 
\begin{equation}
M\propto \left( 
\begin{array}{ccc}
0 & -v_{3} & 0 \\ 
v_{3} & v_{2} & v_{1} \\ 
0 & v_{1} & M_{u}
\end{array}
\right)   \label{eq:15}
\end{equation}
where the scale of the $U(1)$ symmetry breaking is much smaller than the $%
U(2)$ symmetry breaking, \textit{i.e.} $v_{3}\ll v_{1}\simeq v_{2}$. The
zeroes in the entries 
\begin{equation}
M_{13}=M_{31}=M_{11}=0  \label{eq:16}
\end{equation}
are a generic consequence of this class of models: we will exploit the
conditions (\ref{eq:16}) to parameterize the charged lepton mass matrices 
(we make
no assumption on the other entries). Starting from the texture (\ref{eq:16}),
 we will considerably simplify the problem of extracting all mass
hierarchies from the data; nevertheless, this will leave us with a
reasonably large and assorted selection of models. To be more concrete,
after the conditions (\ref{eq:16}) we are left with 6 non-zero entries that
can be parametrized by 6 free variables as follows (an arbitrary phase is 
understood for each entry)
\begin{equation}
M_{\text{lep}}=M_{\text{l}}\left( 
\begin{array}{ccc}
0 & \varepsilon _{1}^{1-p} & 0 \\ 
\varepsilon _{1}^{p} & \varepsilon _{2} & \varepsilon _{2}^{r} \\ 
0 & \varepsilon _{2}^{d} & 1
\end{array}
\right) .    \label{eq:17}
\end{equation}
Neutrinos are described by a Majorana mass matrix that probably arises from 
the seesaw mechanism. The peculiarity of this mechanism  justifies a more 
 prudent approach and a rather universal choice. Thus we prefer the generic mass parameterization: 
\begin{equation}
M_{\text{neutr}}\,/\,\text{eV}=0.06\,\ \left( 
\begin{array}{ccc}
\lambda ^{t_{11}} & \lambda ^{t_{12}} & \lambda ^{t_{13}} \\ 
\lambda ^{t_{12}} & \lambda ^{t_{22}} & \lambda ^{t_{23}} \\ 
\lambda ^{t_{13}} & \lambda ^{t_{23}} & \lambda ^{t_{33}}
\end{array}
\right) .    \label{eq:18}
\end{equation}
where $\lambda =0.2$, while the $t$ exponents are free parameters. Changing 
the constant $\lambda $ or the overall mass scale 0.06 eV, corresponds to a
 linear transormation on the t exponents that has no impact both on the
 a-priori distributions and on the physical results. 
 Clearly one could choose a different reasonable
parameterization with different parameters: nevertheless the needed
parameters would be in one to one correspondence with ours (shown in
(\ref{eq:17},\ref{eq:18})) through well defined equations. Thus we do not loose
in generality, choosing the above parameterization, here the only assumption%
\footnote{%
However this strict equivalence will become only approximately true, after
the implementation of the Monte Carlo procedure in the next section. In fact
a different parameterization would correspond to a different (and non-flat) 
\textit{a-priori} distribution for the exponents $p,r,d$ and $t$.
 This ambiguity is the necessary
prize to pay for our \textit{Bayesian \ }approach.} is equation (\ref{eq:16}).
 The determinant of (\ref{eq:17}) gives the product of the three
eigenvalues and thus $m_{e}m_{\mu }m_{\tau }=\epsilon _{1}M_{l}^{3}.$ We
also observe that $M_{l}\sim m_{\tau }$, and $\epsilon _{2}\sim m_{\mu
}/m_{\tau \mathrm{\ }}$(if $r+d\gtrsim 1$), then we also get that $\epsilon
_{1}\sim m_{e}m_{\mu }/m_{\tau }^{2}.$ The exponents $p,r,d$ and $t$ 
need a more
complex analysis that will be done in the next section.

It is understood that each entry in the (\ref{eq:17},\ref{eq:18}) is
 multiplied by a   coefficient $a_{i}^{l,\nu}$, with  $|a_{i}^{l,\nu}|= 1$
 and 
$0<\mathrm{\arg }(a_{i}^{l,\nu})<2\pi $.

The $U(2)$ symmetry breaking naively discussed above, implies definite
values for the exponents  in the charged lepton sector
\begin{equation}
\begin{array}{ccc}
p=1/2, & r=1, & d=1.
\end{array}
\label{eq:u2}
\end{equation}
However a more sophisticated theoretical study could lead to different
scenarios, as for instance if one embeds the above picture into Grand
Unified Theories. Moreover the above prediction needs to be improved, taking
into account the renormalization group evolution of mass and mixings from
the unification scale  down to the weak scale \cite{rge}.  
In general the predictions of a given model, like the ones of 
eq. (\ref{eq:u2}), 
do not correspond to points in the parameter space but to small balls that 
include the theoretical uncertainties specific of the model.  

We notice that the parametrization in our approach differs slightly from what
 one usually finds in literature. 
In most works the coefficients $a_i$ are assumed to be of order one
 and their incertitudes depend on the model.
In our analyses, to compare different models, we have to fix the modules of
 these coefficients to the values $|a_i| = 1$ and therefore
the model dependent uncertainty of these coefficients is transferred into an
 incertitude on the exponents.
This means that a given model will have not only an intrinsic uncertainty
 on the phases of the coefficients $a_i$ but also a theoretical error on the
 exponents, which depends on the model itself and need to be computed every
 time.
In this philosophy a given model, for example $U(2)$, will not be a point
 but, instead, a {\em ball} in the parameter space.
A model will be more {\em competitive} than another if its {\em ball}
 is in a higher density region. 

Our aim is to
extract these exponents directly from the experimentally measured
quantities, in the same way shown below.

\subsection{Fitting the mass hierarchies}

\textbf{The Method}\newline
The goal is to extract the values of  the exponents $t_{ij}$, $p,r,d$  from the
experimental measurements. A direct fit of the data is not possible, since
the number of free parameters in (\ref{eq:17},\ref{eq:18}) is much larger
than the number of observables, six mass eigenvalues plus the  mixing matrix
parameters.

The main obstacle comes from the coefficients $a_{i}^{l,\nu }$,  whose phases
 are not theoretically known. To cope with them, we will treat
this uncertainty as a theoretical \textit{systematic error}. Namely, we have
assigned a flat probability to all the coefficients $a_{i}^{l,\nu }$ with 
\begin{equation}
0<\arg (a_{i}^{l,\nu })<2\,\pi.  
\end{equation}
The exponents in the charged lepton matrix are chosen in the ranges
\begin{equation}
0<p<1,\,\ \ \ 0<r,d<5.
\end{equation}
$\varepsilon _{1}$ and $\varepsilon _{2}$ are less than one and we randomly
take them with a flat distribution in logarithmic scale. The values of 
$p,r$ and $d$ must satisfy the above constraints since (by definition) we 
choose the entry (3,3) of the matrices in (\ref{eq:17}) to be the largest one.
On the other hand  $t_{ij}$
can take any value: in practice we have chosen an interval $\ -2<t<5$; we
have checked that $t$ is never lower than -2 even if we enlarge the interval. 
For 
$t$ greater than 5, the relative neutrino entry  is close to zero and
negligible. For any random choice of the coefficients $a_{i}^{l,\nu }$, of
the exponents $t,p,r,d$ and the variables $\varepsilon _{1},\varepsilon _{2}$
we get two numerical matrices for, respectively, the charged lepton and the
neutrino sectors. The diagonalization of these matrices gives us six
eigenvalues, corresponding to the physical masses, and two numerical
unitary matrices whose multiplication yields the MNS mixing matrix. We have
collected a large statistical sample of events. Each one of these events can
be compared with the experimental data (see section 2) through a $\chi ^{2}$
analysis: the event is accepted with probability  
\begin{equation}
P(O_{i}^{\mathrm{t}h})\propto{e^{-\frac{1}{2}\chi ^{2}(O_{i}^{\mathrm{t}h})}}
\label{chi2}
\end{equation}
where the $\chi ^{2}$ is defined in (\ref{eq:chisquare}). Before applying
the experimental constraints, the events are homogeneously distributed in the
variables $t,p,r,d,$ and the probability distributions are flat; but after,
applying the weight corresponding to eq. (\ref{chi2}), only points lying in
well defined regions of the space $t,p,r,d$ have a good chance to survive. 
\newline
Let us better clarify the reason for such not uniform distributions and the
physical interpretation of the density of points per unit area in the figures.
 Let us assume two different choices,\footnote{%
We will often use the word ``model'' to understand a particular and fixed
choice of the exponents $p,r,d$ and $t$.} of the exponents $t,p,r$ and $d$
that we call model 1 and model 2, lying in two different regions in the 
 figures; the Monte Carlo generates two samples of lepton and neutrino 
matrices,
through equations (\ref{eq:17},\ref{eq:18}). Only a fraction $p_{1}$ (and $%
p_{2}$) of matrices of the sample 1 (and 2) will pass the experimental
constraints (that is eq. (\ref{chi2})): $p_{1}$ ($p_{2}$) is the probability
that model 1 (2) predicts masses and mixings in a range compatible with the
experiments. The values $p_{1}$ and $p_{2}$ are, respectively, 
proportional to the density of points in region 1 and 2. 
From them we can argued that the model
1 is $p_{1}/p_{2}$ more (or less, if $p_{1}/p_{2}<1$ ) likely than model 2.
Even if our Monte Carlo approach favours most predictive and accurate
models, we also emphasize that one should not mistake these results with
true experimental measurements. They only give us ``natural'' range of
values for the exponents $t,p,r$ and $d$. \newline
If $p_{max}$ is the density of points at the maximum, we call $R=p/p_{max}$
the ratio of the probability with respect the value at the maximum, in each 
figure. By
definition $R \leq 1$. We will show the regions corresponding to different values
of R.
\subsection{Results}
In all figures  points are blue (dark) in regions where $R>0.1$, red when 
 $0.05<R<0.1$ and green (light) when $R<0.05$.
We  show eighteen figures corresponding to different pairings of the
 exponents both in  the charged lepton and in the neutrino sector.

A comment is in order here. In  any large set of  experimental measurements, 
the fact that   one measurement  
 deviates from the  theoretical prediction by two standard deviations,  should
 not be seen as  a problem: if the set of data is large, it is expected
 that few  measurements  slightly depart from the theoretical expectations.
 Similarly, if we take a model with definite exponents $t$, it could fall 
in the  green region in one of the shown figures. This should not be
 interpreted as problematic.  But  if the same model 
 lives in the green region of  two or more   figures,
 than one should start worrying  about  its  naturalness.

\subsubsection{Charged lepton sector and SU(5) Unification}

Let us start from the figures for the charged lepton sector. An interesting
result can be deduced from the figure $p$ vs $r$. It is clear that a value
for $r$ close to zero ($r\lesssim 0.2$) is favored by data. This reminds us
an analogous result obtained in the quark sector. In that case \cite{stocchi}%
, it was the exponent $d$ in a parameterization of the down quark sector
similar to the (\ref{eq:17}). This coincidence can be considered as a hint
for an $SU(5)$ grand unification. In fact, if we look at the fermion mass
matrix in $SU(5)$, both charged leptons and down quark mass matrices arise
from the same Yukawa interaction. Namely if $T_{i}^{ab}$ and $F_{i}^{b}$
represent the $10$ and $\bar{5}$ of the generation $i$, than the Yukawa
interactions (in a $SU(5)$ unified model)
\begin{equation}
g_{ij}\,\ T_{i}^{ab}\,F_{j}^{b}\,H^{a}  \label{yuk}
\end{equation}
simultaneously give the two mass matrices for the down quarks and the
charged leptons. Since $\,F_{j}^{b}$ contains the right-handed quarks and
the left-handed leptons, the two matrices generated by the (\ref{yuk}), are
one the transposition matrix of the other, (i.e. with left and
right indices exchanged). An anomalously large $g_{32}\,\ \ (\sim $ $%
g_{33})$ in (\ref{yuk}), will simultaneoulsy give $d_{\text{down}}\lesssim
0.2$ in the down sector and $r_{\text{lepton}}\lesssim 0.2$ in the charged
lepton sector. One can also note that $0.2<p<0.7$, namely strongly
asymmetric textures in (1,2) sector of charged leptons is not ruled out. 
Moreover from the figure $p$ vs $d$ we have no evidence that a symmetric texture is preferred,
and numerically we could also have $(M_{\text{lep}})_{12}/M_{\text{l}}\sim $ 
$(M_{\text{lep}})_{21}^{4}/M_{\text{l}}^{4}$ or $(M_{\text{lep}})_{21}/M_{%
\text{l}}\sim $ $(M_{\text{lep}})_{12}^{4}/M_{\text{l}}^{4}$.

\subsubsection*{Neutrino masses}

Concerning the neutrino mass texture (\ref{eq:18}) we can plot 15 figures \
corresponding to all different pairings of the 6 independent exponents in
the parameterization (\ref{eq:18}). There are two main features that can be
immediately observed. Plots that differ for a simple exchange of the label 2
with 3 are identical. For instance the plot 12 vs 33 is identical to 13 vs
22. This result is not at all trivial: in the charged lepton sector we have
found that $g_{32}\sim $ $g_{33}$, \textit{i.e.} both entries are of order
1; neglecting all other small entries, such a matrix for the charged lepton
shows an approximate discrete symmetry that exchanges the 2nd and 3rd \
rows, namely the left-handed components of the  leptons of the second and
third families. We remind that  by definition of the 
parameterization (\ref{eq:18}), 2 and 3 labels the neutrino states that form a 
$SU(2)$ doublet
with the charged leptons of respectively 2nd and 3rd generation. Thus the
above symmetry implies that for any  neutrino mass matrix, fitting the
data, there must be another solution with 2 and 3 exchanged.

From a preliminary inspection of the full set of neutrino figures we also
remark that some plots provide us with more stringent constraints than
others. This make clear which theoretical ingredient is essential (and
 which is not), to describe the neutrino phenomenology. 
The plots 22 vs 33 , 23 vs 33,  23 vs 22, 12 vs 13 , 12 vs 22   put
significant constraints, while 12 vs 11 or 13 vs 11 are statistically less
significant (since points are spread over a wider region). We can \
deduce a lower bound for the exponent $t_{11}\gtrsim 1$.  Other pictures
( 22 vs 11, 13 vs 11, 23 vs 11)  show that  $t_{11}$ is not
correlated with other entries, thus  this lower bound holds regardless
of other entries. 

The most clean figure is 22 vs 33. As already mentioned,
the allowed region is symmetric under the exchange $22\leftrightarrow 33$.
Even if $t_{22}$ can take any value from zero to infinity, (\textit{i.e. } $%
m_{22}=0.06\,\lambda ^{t_{22}}<0.06$ eV), $t_{22}$ is very strongly
correlated to $t_{33},$ and  a rather precise value for $t_{22}\,$ can be
deduced once $t_{33}$ (or equivalently $m_{33}$) is fixed. We can
distinguish two cases: one when $t_{22}>>t_{33}\sim 0$ (or when $%
t_{33}>>t_{22}$) and another with $t_{22}\sim t_{33}\sim 0.5$. In the first
 case $%
t_{33}\sim 0$ and the mass entry $m_{33}^{2}\sim 0.0036\,\,\ eV^{2}\sim \Delta
m_{32}^{2}$ explains the atmospheric neutrino mass, while $t_{22}$ is
preferably $1\lesssim t_{22}\lesssim 2$. Then from figure 12 vs 22 we learn
that also $0.5 \lesssim t_{12}\lesssim 2$. From fig $\ $\ 12 vs 13 we learn
that $t_{13}\gtrsim 1.$. This plot shows an important correlation
between $t_{12}$ and $t_{13}$, and at least one of them must be close to
1 - 1.5.  Finally, exploiting 23 vs 22 and 23 vs 33 we can set a lower limit $%
t_{23}\gtrsim 0.5$. Only when $t_{22}$ or $t_{33}$ falls in the
interval \ $\ 0.5<t_{22},t_{33}<1$ or $t_{22},t_{33}>2,$ then we can also
give an upper limit $t_{23}\lesssim 1.$

In addition to this generic conclusion we report in table 1 some regions
that maximize the probability distribution, and that result as most likely,
 from our study.
\subsubsection{Neutrino Anarchy}

Another possible scenario emerges when $t_{22}\sim t_{33}\sim 0.5$ (see
fig. 22 vs 33). In such a case fig. 23 vs 22 (and 23 vs 33) tells us that
also $t_{23}\sim 0.5$. This scenario reminds us  the so-called neutrino
anarchy~\cite{anarchy},  however figures 12 vs 13 and 12 vs 11 clearly
forbid value as low as 0.5 for $t_{11},t_{12}$ and $t_{13}$.  Namely, while
the heaviest squared $2\times 2$ sub-matrix  elements are of the same order
of magnitude ($\sim 0.03$ eV),  the  first row and column of the neutrino
mass matrix must be smaller by at least a factor $\sim \lambda ^{0.5}=0.4,$
with respect to the heaviest entries. This result is compatible with the
particular scenario studied in~\cite{Vissani:2001im}
and disfavors strict neutrino anarchy.

\begin{table}[tbp] \centering%
%
{\ }${
\begin{tabular}{|ccc|}
\hline
$2.<t_{13}$ & $2.<{t_{23}}$ & $0.7<t_{12}<1.8$ \\ \hline
$-0.13<{t_{33}}<0.13$ & ${1<t_{11}}$ & $1<{t_{22}}<2.$ \\ \hline\\\hline
$1.<t_{13}<2.$ & $2.<{t_{23}}$ & $0.6<t_{12}<2.$ \\ \hline 
$-0.13<{t_{33}}<0.13$ & ${1<t_{11}}$ & $1<{t_{22}}<2.$ \\ \hline\\ \hline
$2.<t_{13}$ & $0.5<{t_{23}}<2.$ & $0.6<t_{12}<2.$ \\ \hline
$-0.15<{t_{33}}<0.2$ & ${1<t_{11}}$ & $1<{t_{22}}<2.$ \\ \hline
\end{tabular}
}$\caption{As discussed in the text, we present the three regions in the 
parameter space that are the most likely in our analysis. The probability 
corresponding to the first one is about three times larger than the one 
corresponding to the other two.
We found with the same probability also three other regions that can be 
obtained from these ones by a full exchange of the index 2 with the index 3 
in the $t_{ij}$ values. \label{key}}%
\end{table}%
%
{\ }

\subsubsection{Neutrinoless Double Beta Decay}

As said in the second section, the set of data we used includes the 
neutrinoless double beta
decay result. However, we stress that this data has a negligible impact to
all shown figures. In fact, we have found that in all selected regions  the
corresponding $m_{ee}$ is very small and incompatible with the neutrinoless
double $\beta $ decay (N$2\beta$D) experiment. There is only a tiny region where $%
m_{ee}$ starts being sizable, and it can be clearly seen in figures 12 vs
33, 13 vs 22  but also  in 11 vs 33 , 11 vs 23  and 11 vs 22: few points
fall in regions with negative values for the exponent $t$.
 These points are very distinct, since
they  form a thin line rather than a scattered region. When $t$ is
negative, the corresponding mass entry is larger than 0.06: on one hand this
large value allows larger values of $m_{ee}$, but  on the other hand the
relative small experimental value $\Delta m_{32}^{2}$
  requires a precise tuning of the physical masses and the exponents.

Thus we conclude that our analysis disfavours\footnote{%
However one could give a more optimistic interpretation to our result: if
the experimental debate on the evidence for a neutrinoless double beta decay
 would confirm this evidence, the theoretical impact would be dramatic.
Almost all regions in our plots would be ruled out, and we were left with
few points, and very clean and distinct solutions for the neutrino mass
matrix.}  a large $m_{ee}$ , at least as high as needed to explain N$2\beta$D.

\section{Conclusions}

We can summarize our results as follow. There is a hint for an
asymmetric non trivial texture in the charge lepton sector. 
Combined with an analogous
result in the down quark mass matrix, this can be interpreted as a new
hint of $SU(5)$ grand unification. In addition to the well know bottom-tau
unification, the entries 32 of the down quark and charged
lepton mass matrices seem to unify. Given also that $m_{s}/m_{b}\sim m_{\mu }/m_{\tau }$
we can conclude that $SU(5)$ unification works sufficiently well in the $%
2\times 2$ sub-matrix formed with the second and third generations. In other
words, not only the physical mass unify but also the large mixing angle
between the  right-handed strange and the bottom  unifies with the large
mixing angle between the left-handed muon and tau leptons. The latter is
needed to explain the large atmospheric neutrino mixing angle and the former
can better fit the CKM parameters \cite{stocchi}. This left-right 
asymmetric texture  suggests that the \textit{naive }%
$U(2)$  must be improved: while the 10 of $SU(5)$  can succesfully transform as
a doublet under $U(2)$,  the same choice for the $ \bar 5$ seems disfavoured by
data. Taking the $\bar 5$  as singlets under $U(2)$, can better accomodate the
large left-right asymmetry. This choice is also supported by the neutrino
data. The left-handed neutrino components belongs to $F$, thus if these were
doublet under $U(2)$ we would expect a relatively large hierarchy  among
different neutrino matrix elements.

\subsection{Acknowledments}
We are really grateful to E. Torrente-Lujan and to P. Aliani for many 
enlightening discussions about neutrino physics.
MIUR is acknowledged for the financial support.

\begin{figure}[cc]
\vskip  0.6cm
\epsfig{figure= 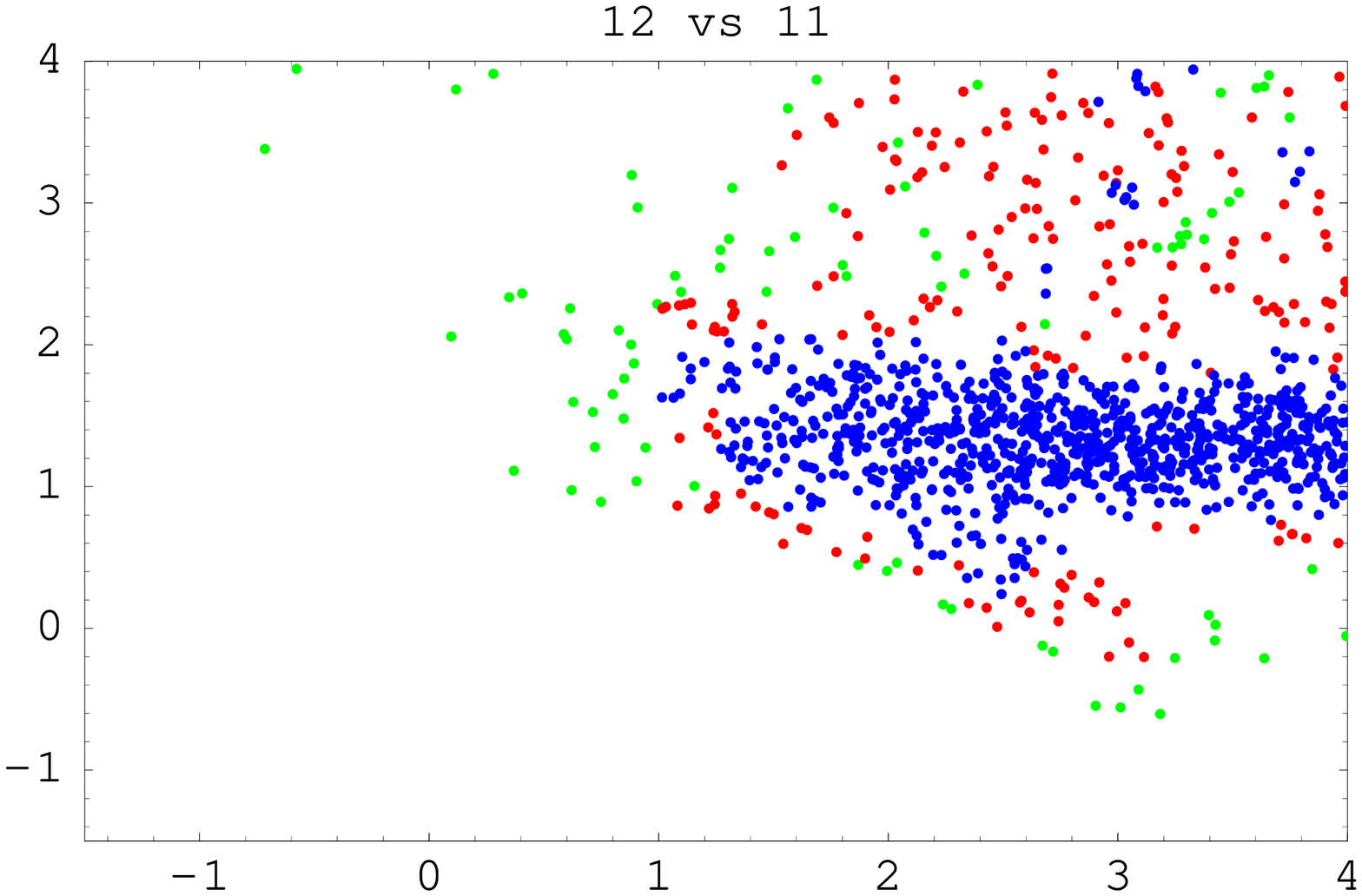, height= 6cm}
\epsfig{figure= 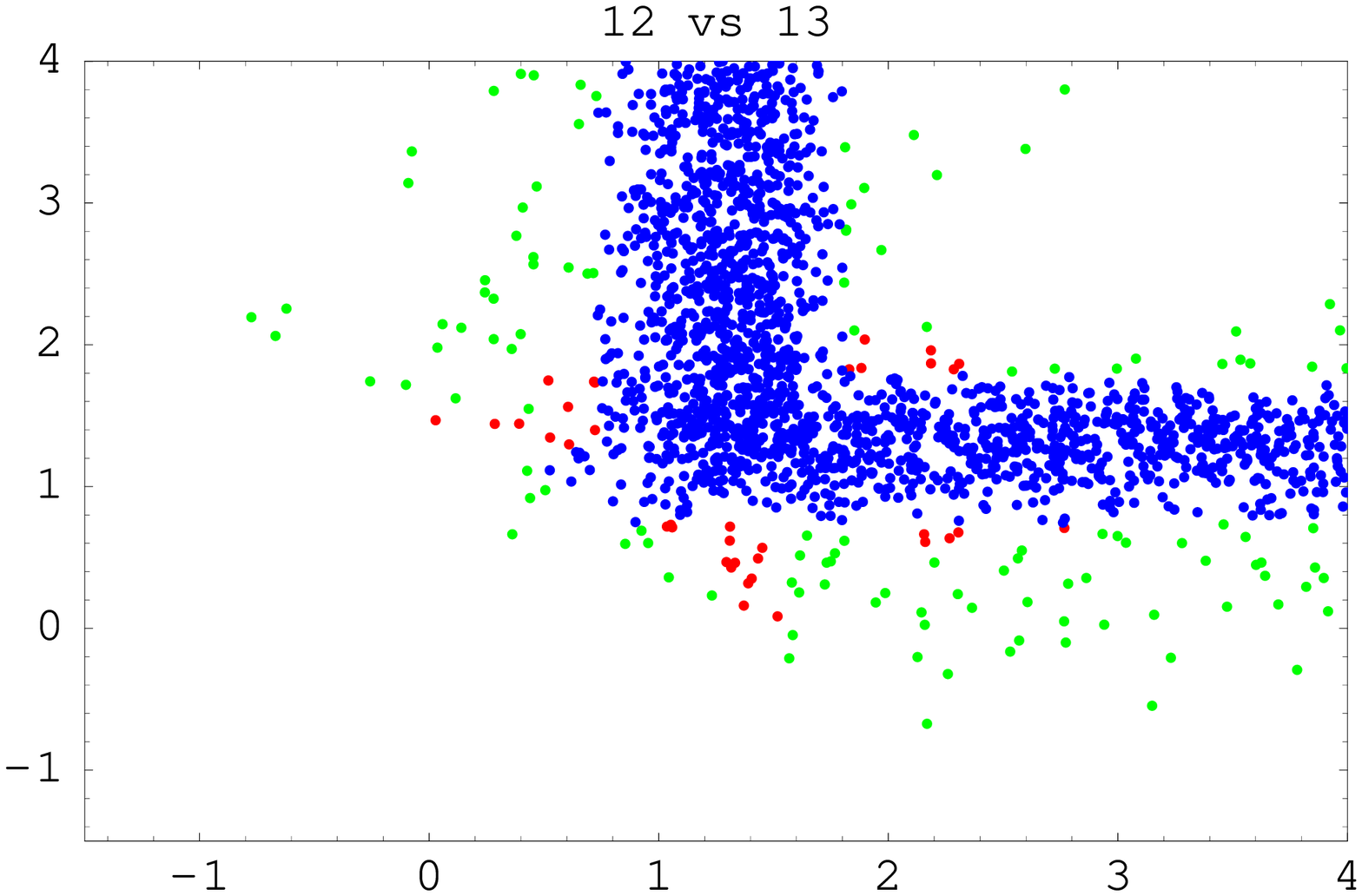, height= 6cm}
\epsfig{figure= 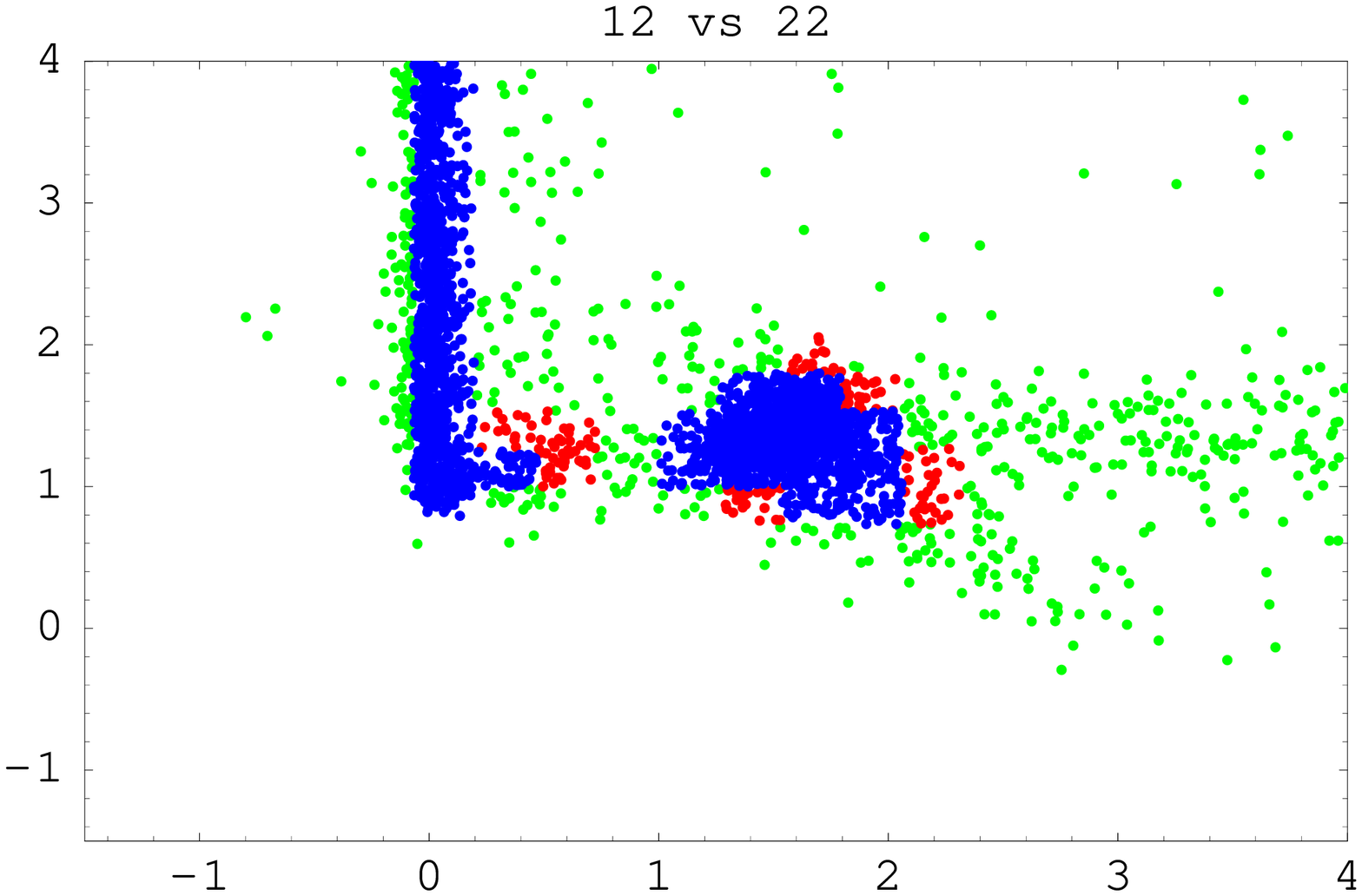, height= 6cm}
\caption{The exponent t$_{12}$ (vertical axis) versus t$_{11}$, t$_{13}$ and 
t$_{22}$. }
\end{figure}
\begin{figure}[cc]
\vskip  0.6cm
\epsfig{figure= 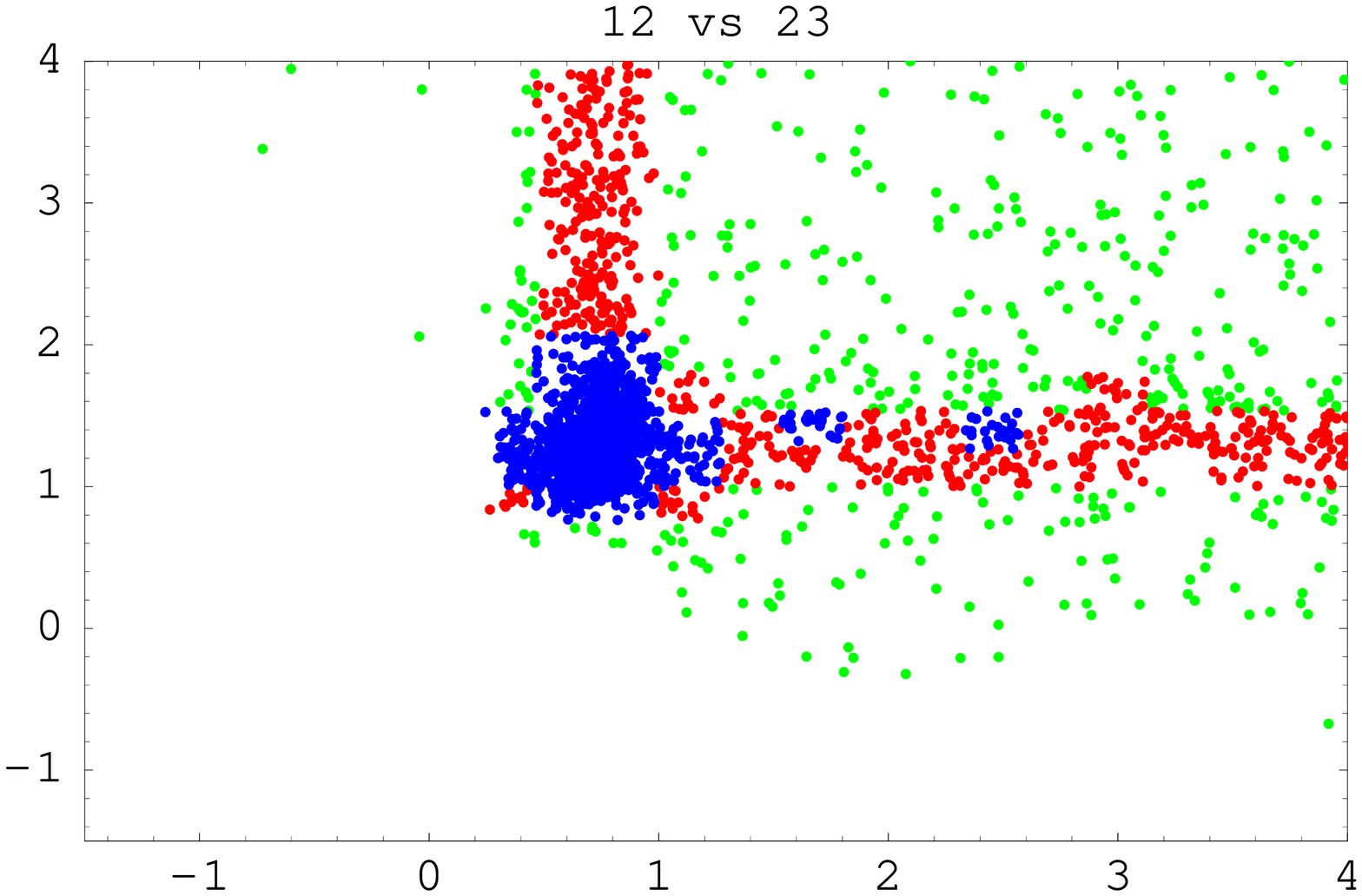, height= 6cm}
\epsfig{figure= 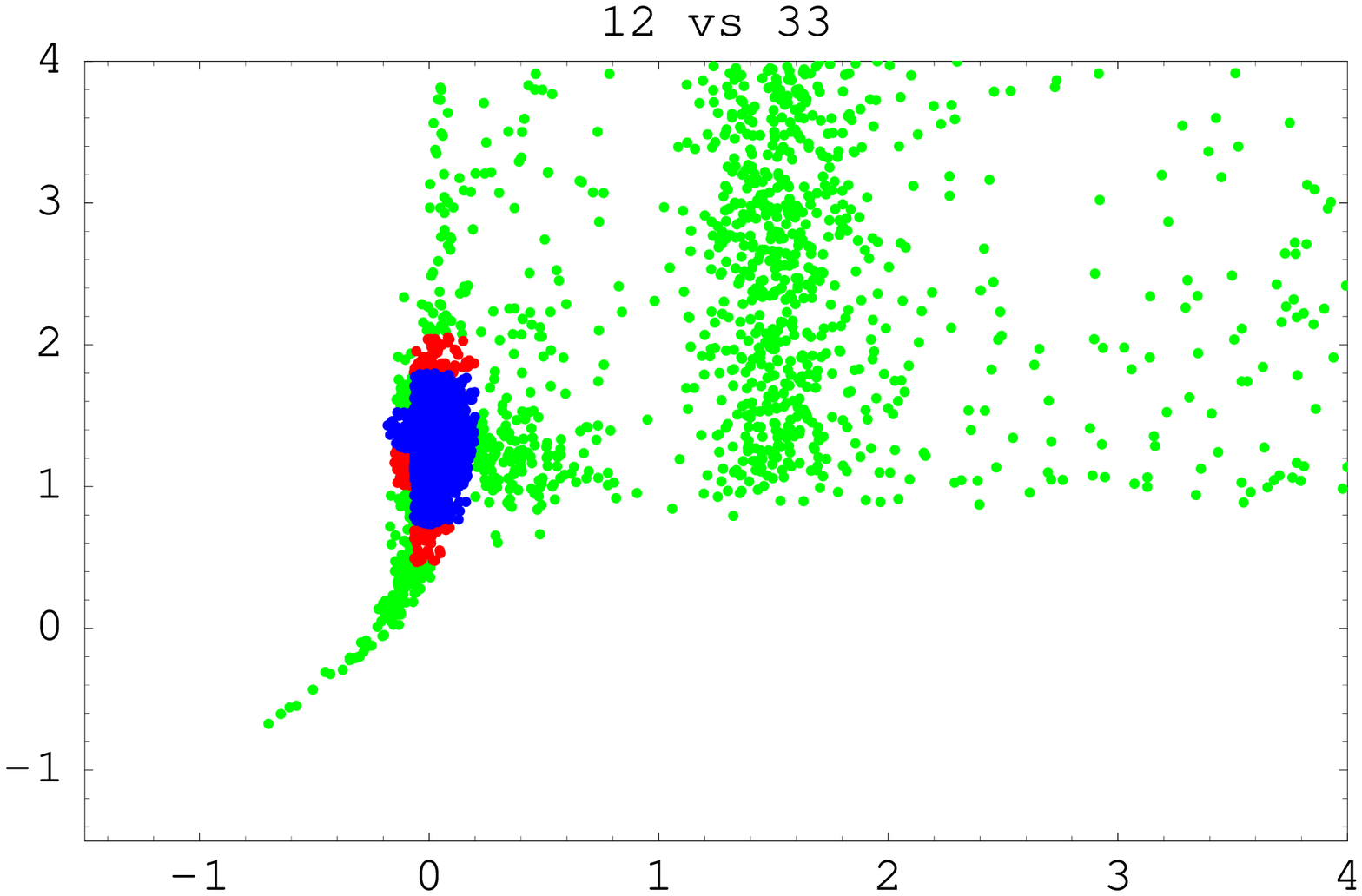, height= 6cm}
\epsfig{figure= 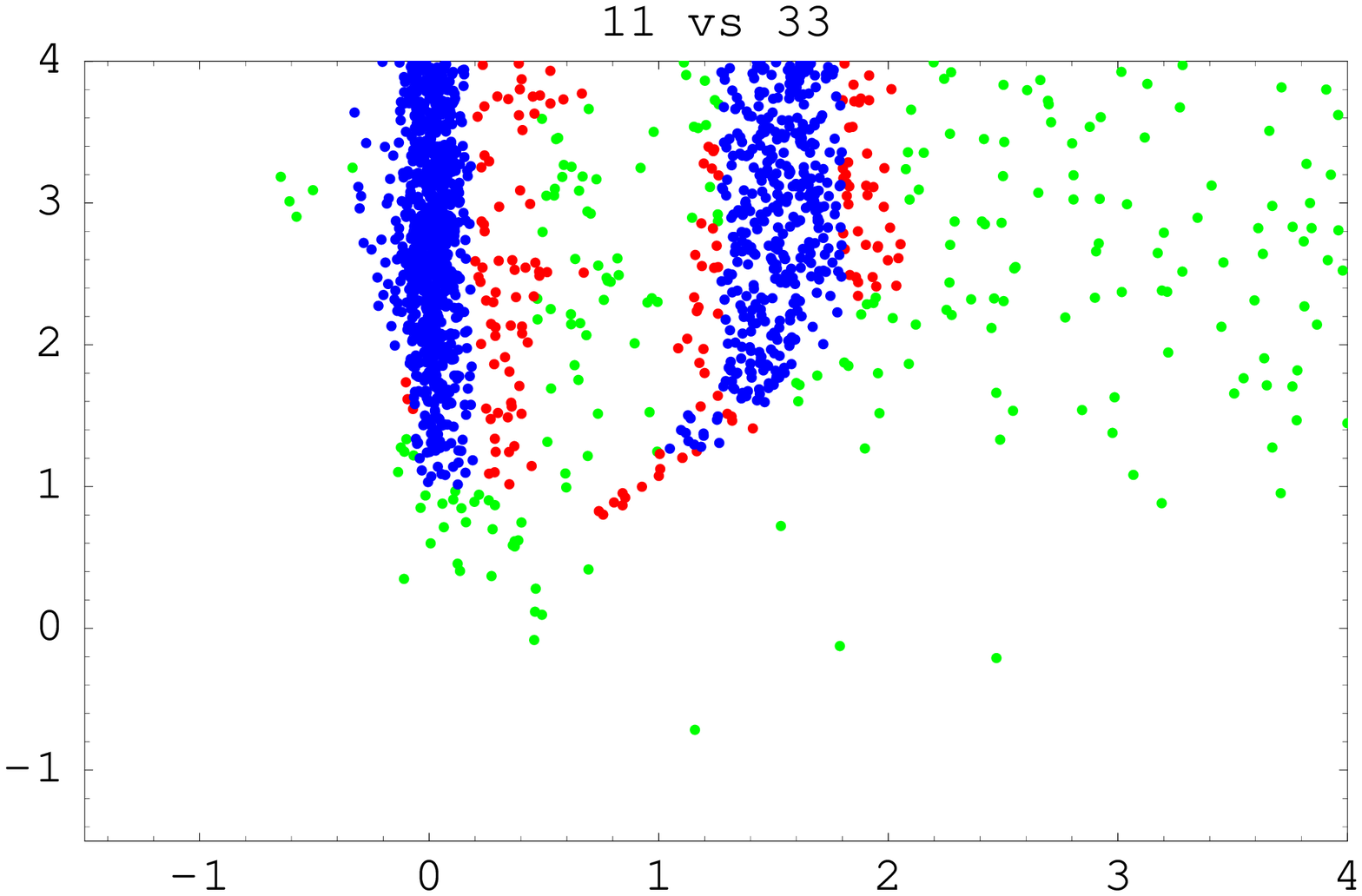, height= 6cm}
\caption{The exponent $t_{12}$ (vertical axis) versus $t_{23}$, $t_{33}$ and 
in the third figure the exponent $t_{11}$ (vertical axis) versus $t_{33}$. }
\end{figure}
\begin{figure}[cc]
\vskip  0.6cm
\epsfig{figure= 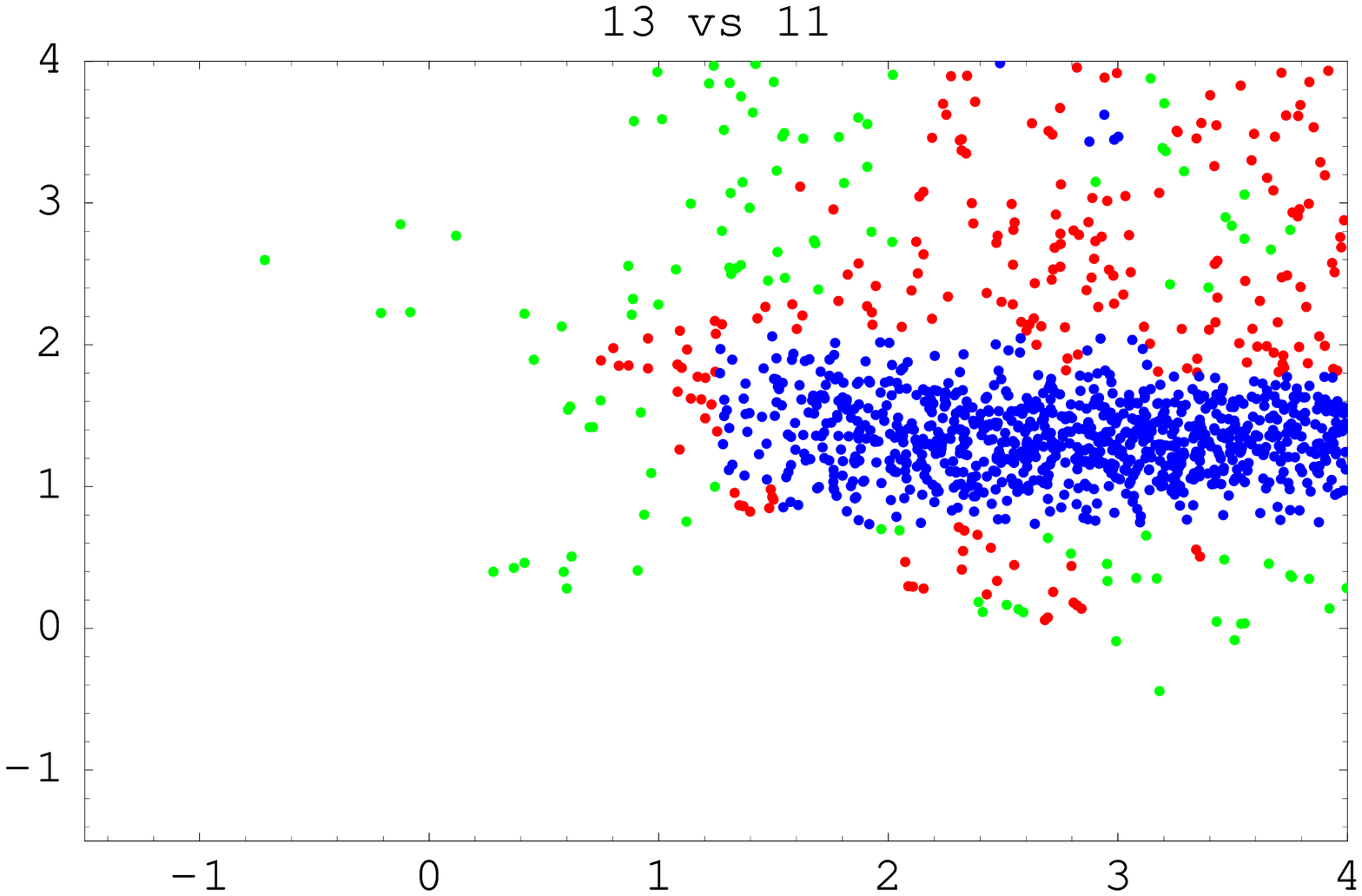, height= 6cm}
\epsfig{figure= 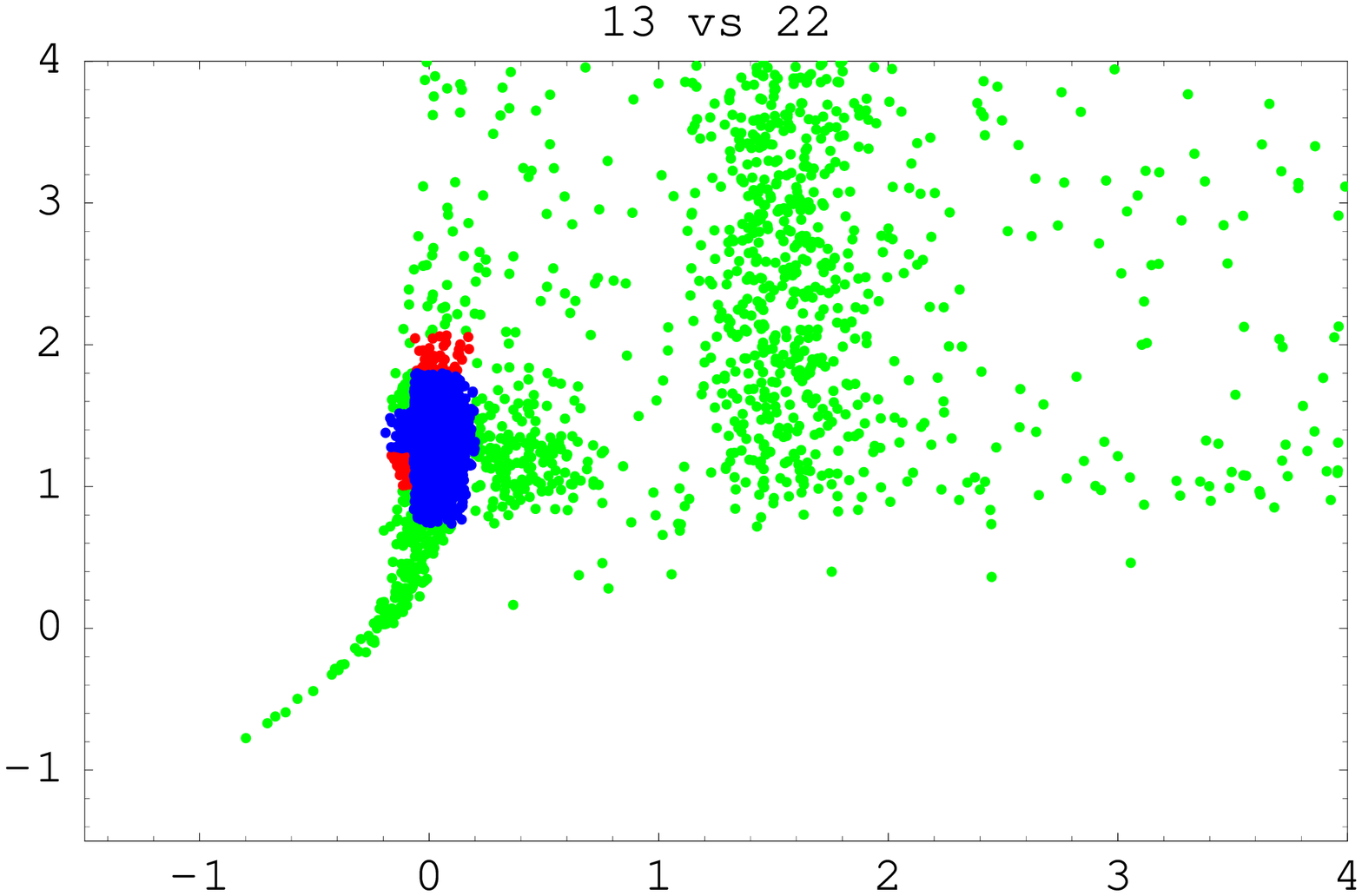, height= 6cm}
\epsfig{figure= 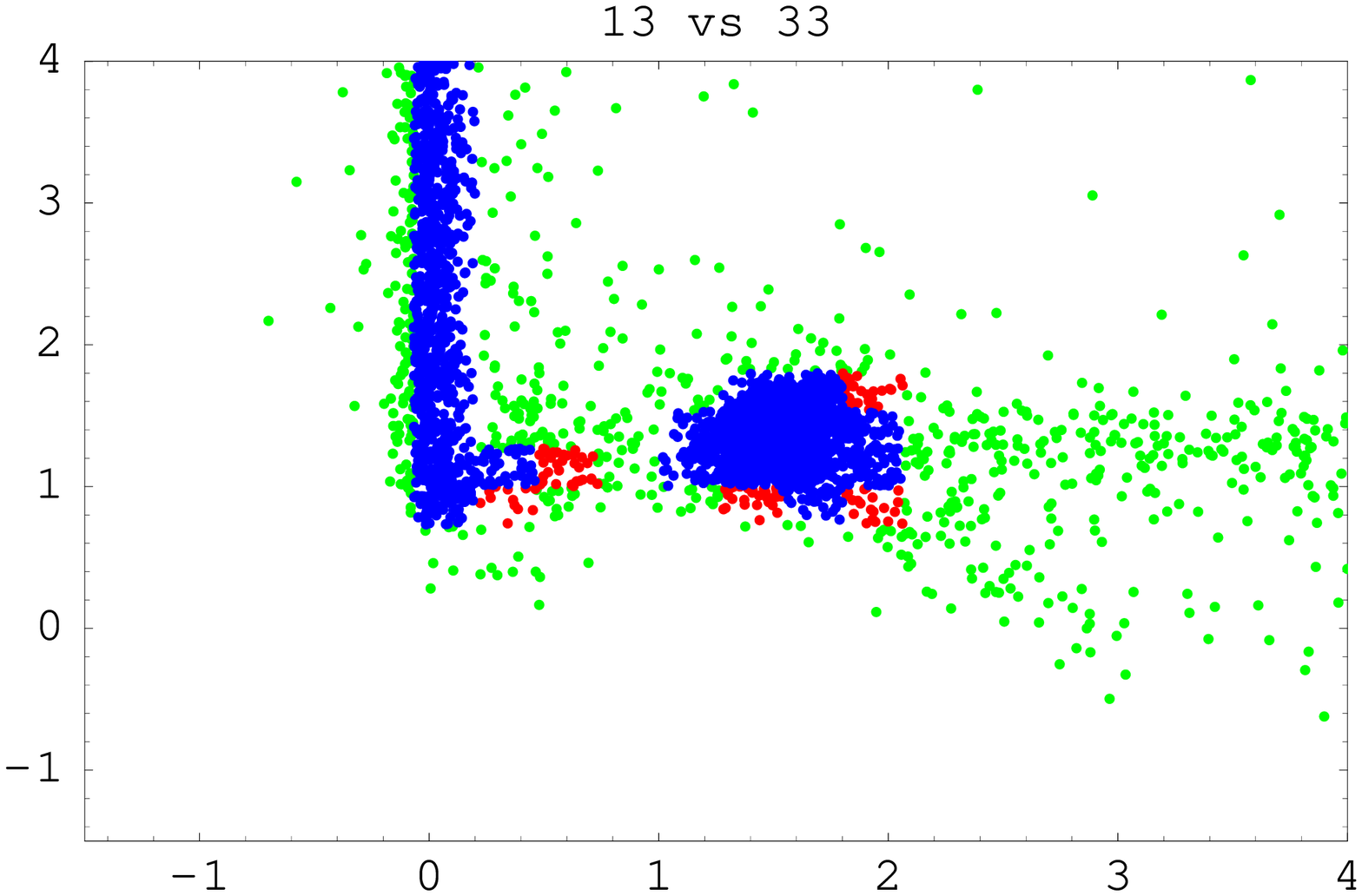, height= 6cm}
\caption{The exponent $t_{13}$ (vertical axis) vs $t_{11}$, $t_{22}$ and 
$t_{33}$. }
\end{figure}
\begin{figure}[cc]
\vskip  0.6cm
\epsfig{figure= 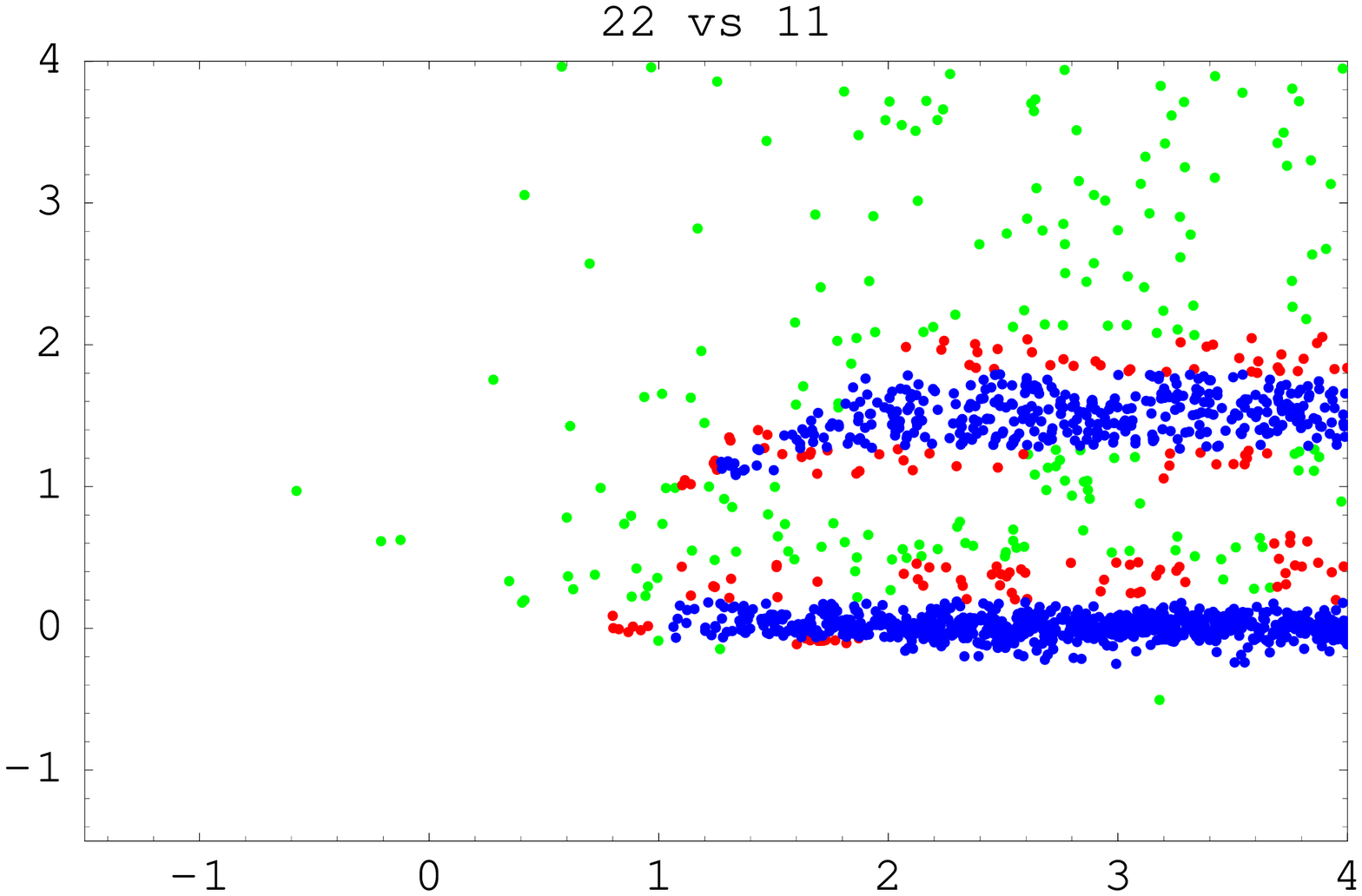, height= 6cm}
\epsfig{figure= 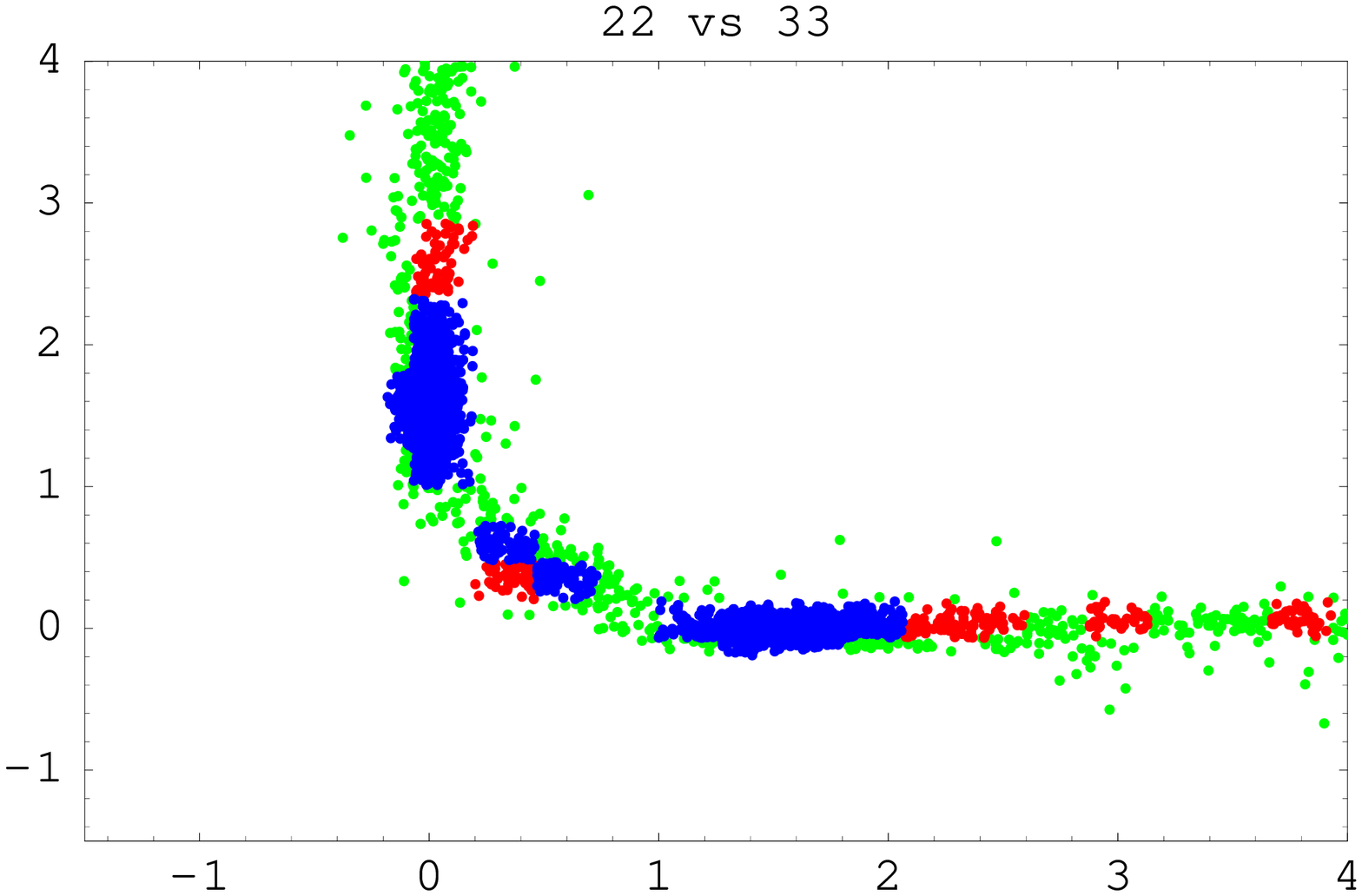, height= 6cm}
\epsfig{figure= 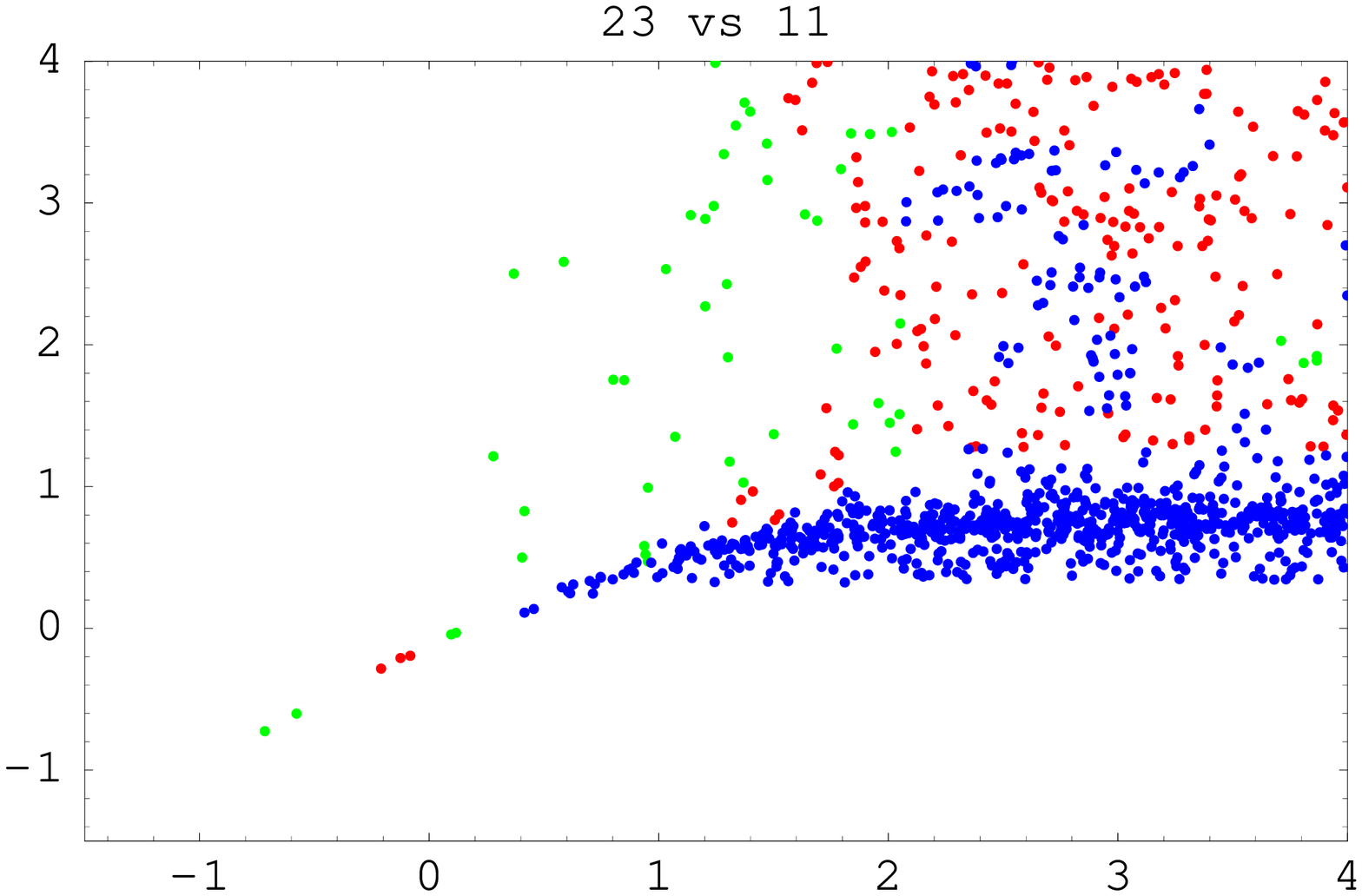, height= 6cm}
\caption{The exponent t$_{22}$ (vertical axis) vs $t_{11}$, $t_{33}$ and 
in the last figure the exponent $t_{23}$ vs $t_{11}$. }
\end{figure}
\begin{figure}[cc]
\vskip  0.6cm
\epsfig{figure= 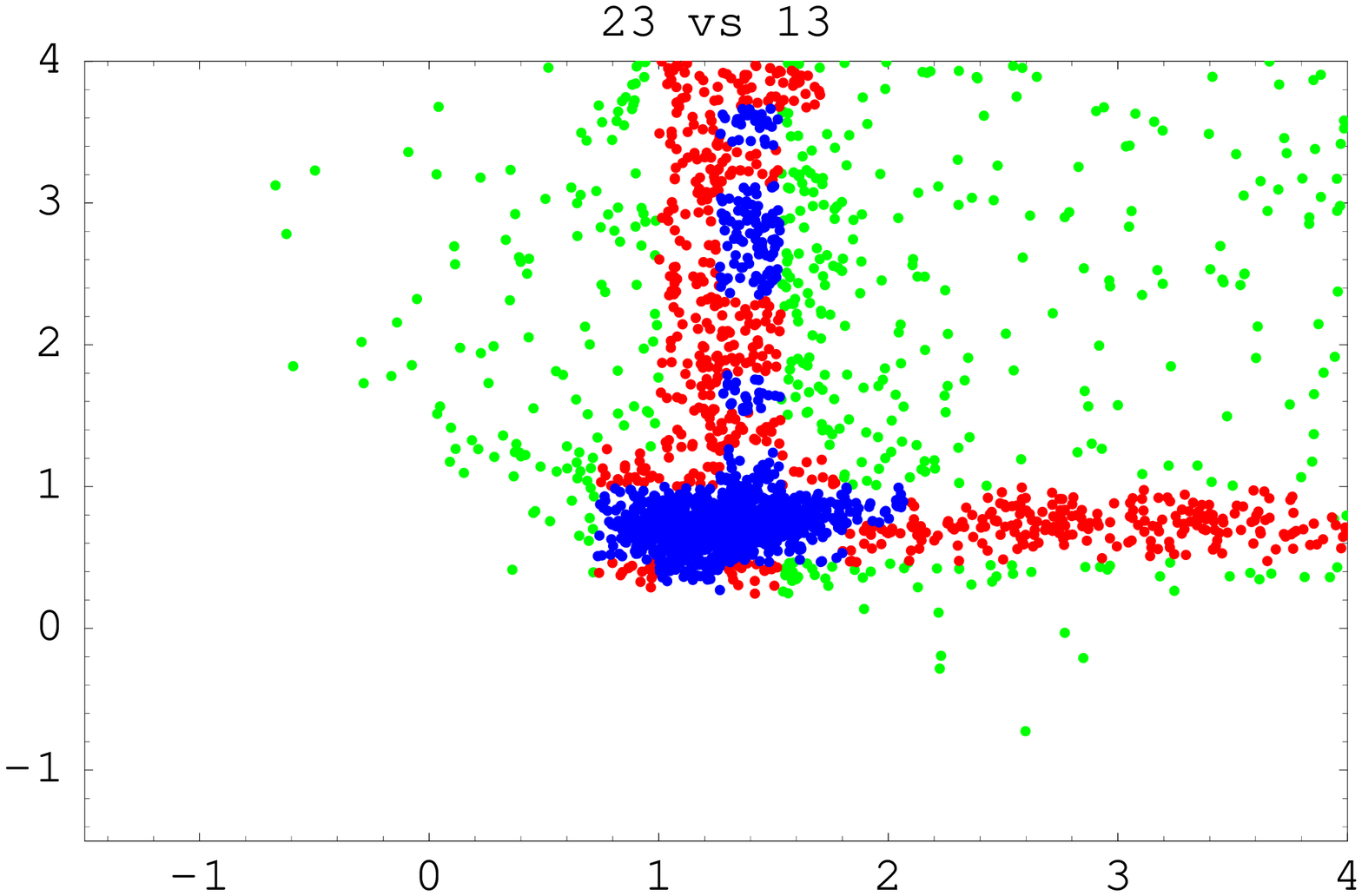, height= 6cm}
\epsfig{figure= 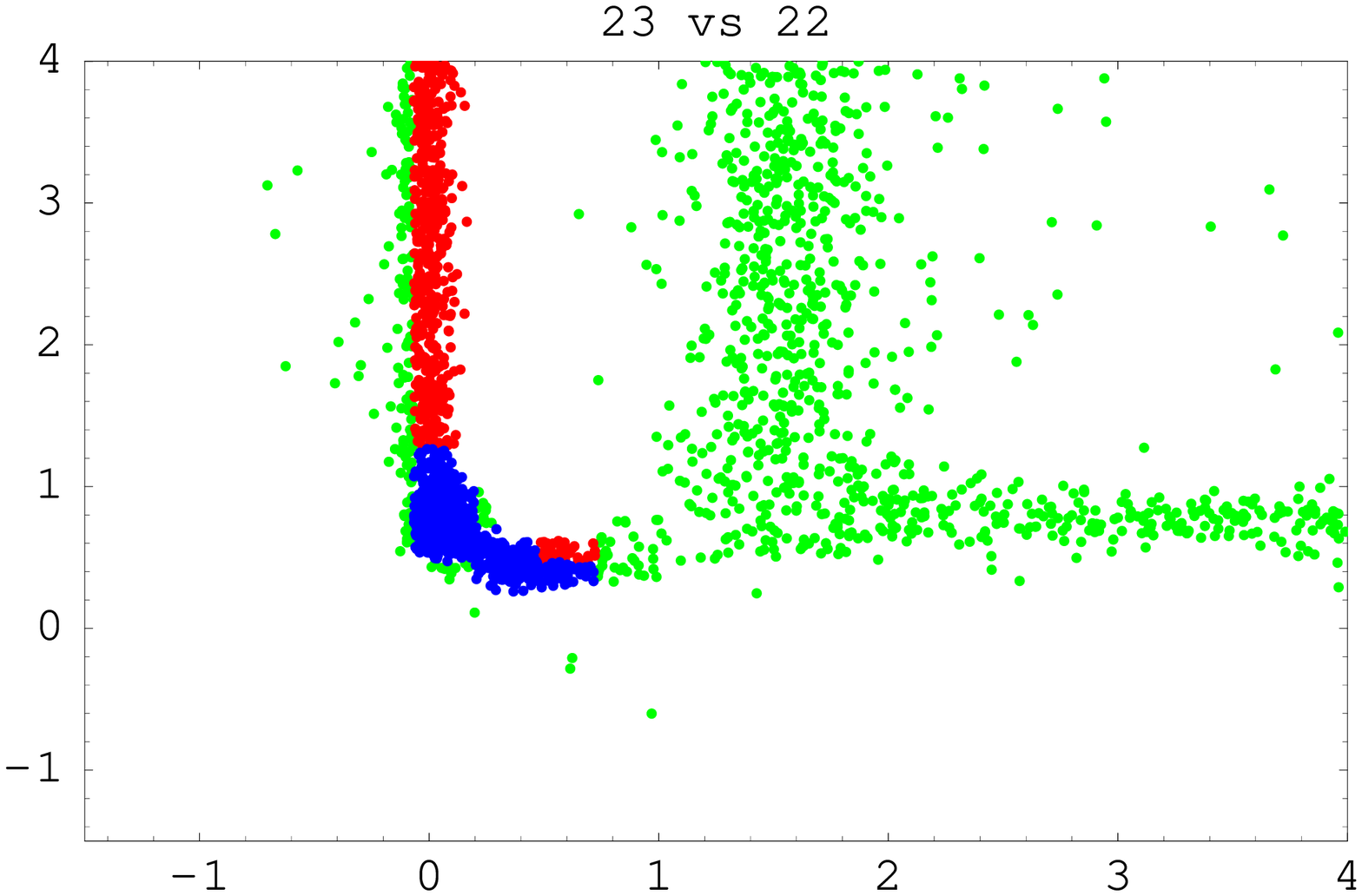, height= 6cm}
\epsfig{figure= 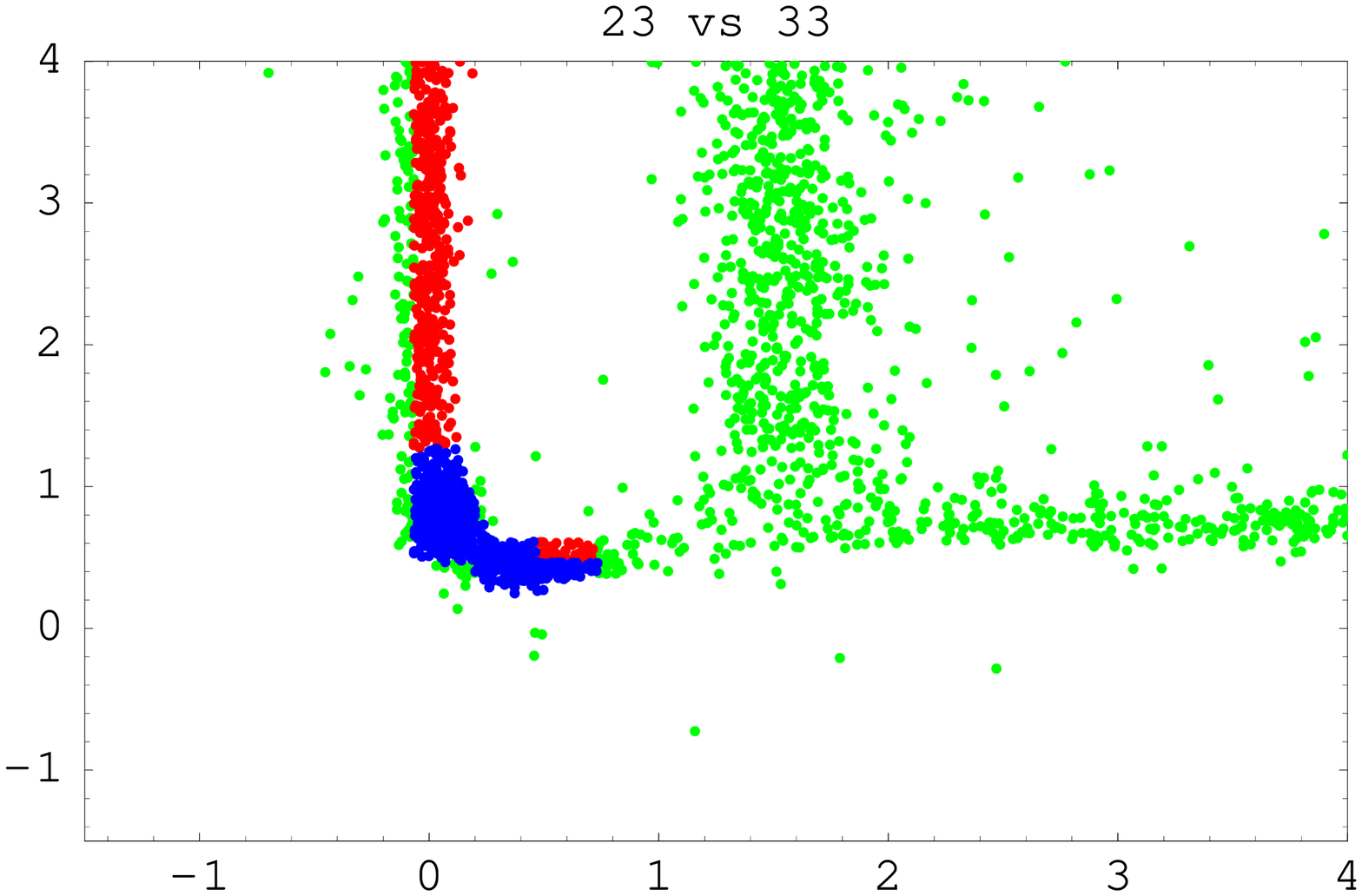, height= 6cm}
\caption{The exponent $t_{23}$ (vertical axis) vs $t_{13}$, $t_{22}$ and $t_{33}$. }
\end{figure}
\vfill 
\eject

\begin{figure}[cc]
\vskip  0.6cm
\epsfig{figure= 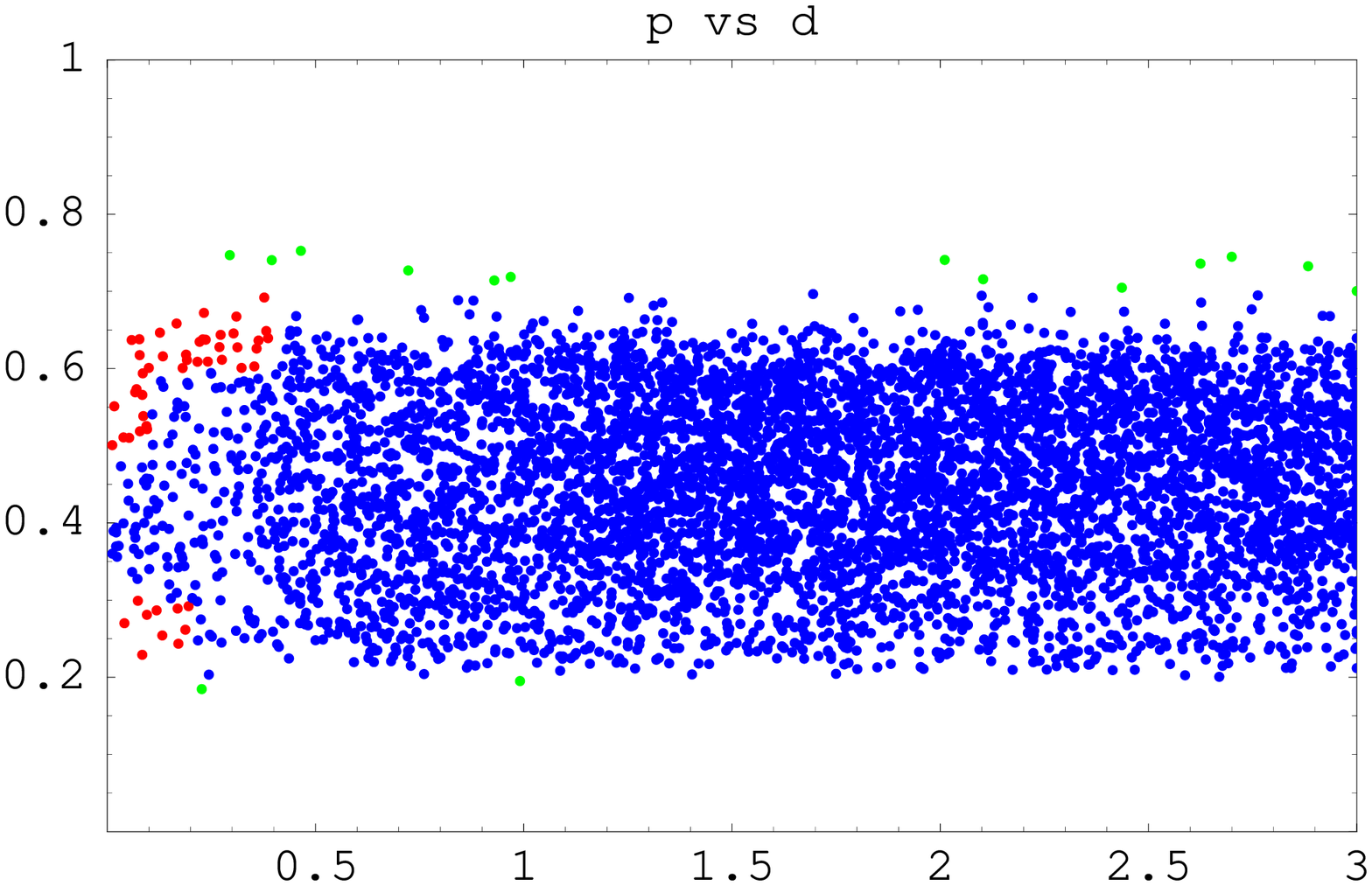, height= 6cm}
\epsfig{figure= 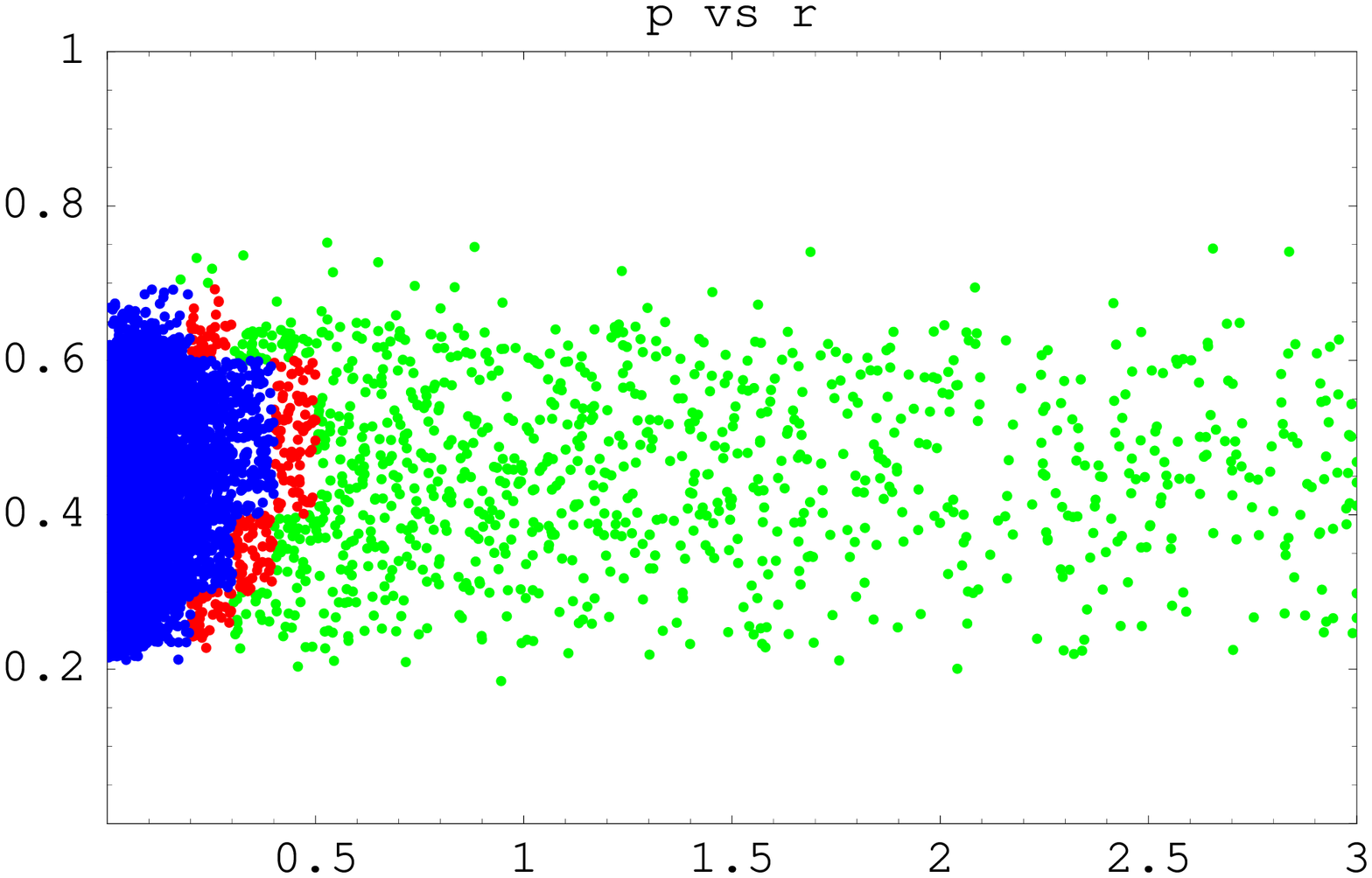, height= 6cm}
\epsfig{figure= 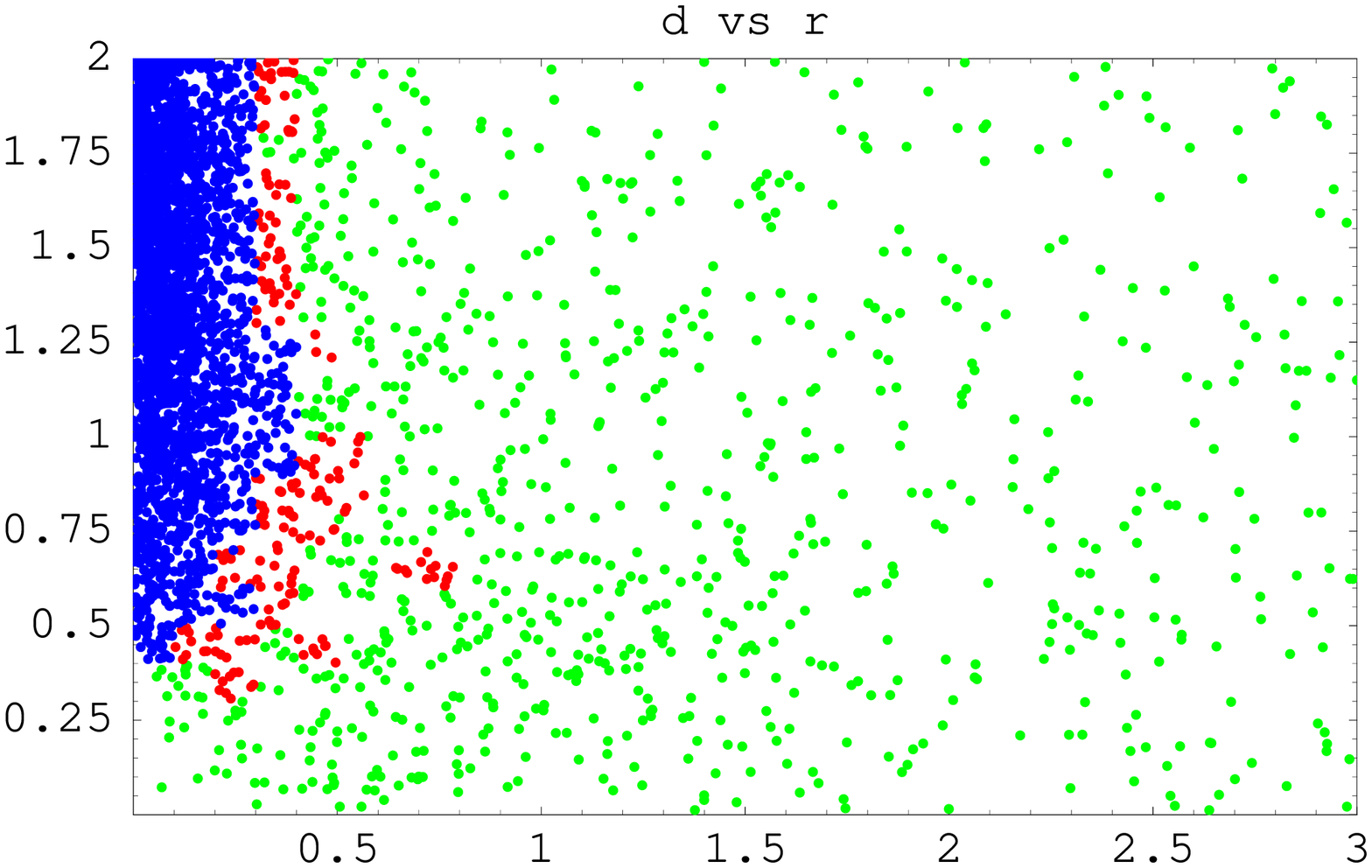, height= 6cm}
\caption{The exponent $p$ (vertical axis) vs $d$ and $r$ and, in the last 
figure, the exponent d (vertical axis) vs $r$. }
\end{figure}
\vfill
\eject

\end{document}